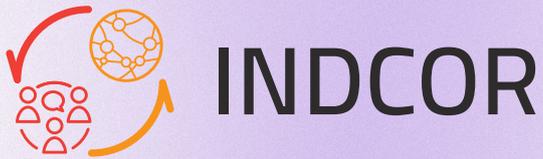

# INDCOR

INDCOR WHITEPAPER 5

# Addressing Societal Issues in Interactive Digital Narratives

**AUTHORS**
Cláudia Silva, Juan Miguel Aguado, Dren Gërguri, Ledia Kazazi, Björn Berg Marklund, Rocío Zamora Medina, Shahira S. Fahmy, José Manuel Noguera Vivo, Eliane Bettocchi, Tao Papaioannou, Maitê Gil, Lissa Holloway-Attaway, Hartmut Koenitz

cost
EUROPEAN COOPERATION
IN SCIENCE & TECHNOLOGY

# INDCOR
**AUTHORS**

Cláudia Silva, Juan Miguel Aguado, Dren Gërguri, Ledia Kazazi, Björn Berg Marklund, Rocío Zamora Medina, Shahira S. Fahmy, José Manuel Noguera Vivo, Eliane Bettocchi, Tao Papaioannou, Maitê Gil, Lissa Holloway-Attaway, Hartmut Koenitz

INDCOR WHITEPAPER 5

# Addressing Societal Issues in Interactive Digital Narratives

This article/publication is based upon work from COST Action INDCOR, CA18230, supported by COST (European Cooperation in Science and Technology).

COST (European Cooperation in Science and Technology) is a funding agency for research and innovation networks. Our Actions help connect research initiatives across Europe and enable scientists to grow their ideas by sharing them with their peers. This boosts their research, career and innovation.

www.cost.eu

# Table of Contents

# Executive Summary

This white paper introduces Interactive Digital Narratives (IDN) as a powerful tool for tackling the complex challenges we face in today's society. In the scope of COST Action 18230 - Interactive Narrative Design for Complexity Representation (INDCOR), a group of researchers dedicated to studying media selected five case studies of IDNs, including educational games and news media, that confront and challenge the existing traditional media landscape. These case studies cover a wide range of important societal issues, such as racism, coloniality, cultural heritage, war, and disinformation. By exploring this broad range of examples, we aim to demonstrate how IDN can effectively address social complexity in an interactive, participatory, and engaging manner. We encourage you to examine these cases and discover for yourself how IDN can be used as a creative tool to address complex societal issues.

This white paper might be inspiring for journalists, digital content creators, game designers, developers, educators using information and communication technologies in the classroom, or anyone interested in learning how to use IDN tools to tackle complex societal issues. In this sense, along with key scientific references, we offer key takeaways at the end of this white paper that might be helpful for media practitioners at large, in two main ways: 1) Designing IDNs to address complex societal issues and 2) Using IDNs to engage audiences with complex societal issues.

# Related INDCOR Whitepapers

**INDCOR WP 0: Interactive Digital Narratives (IDNs) - A solution to the challenge of representing complex issues**
https://arxiv.org/abs/2306.17498
Provides an introduction to Interactive Digital Narratives and representations of complexity

---

**INDCOR WP 1: A Shared Vocabulary for IDNs**
https://arxiv.org/abs/2010.10135
Describes INDCOR's effort to create a shared vocabulary for creators, scholars and analysts

---

**INDCOR WP 2: Interactive Narrative Design for Representing Complexity**
https://doi.org/10.48550/arXiv.2305.01925
Describes Design approaches for IDN

---

**INDCOR WP 3: Interactive Digital Narratives and Interaction**
https://arxiv.org/abs/2306.10547
Describes the concepts behind IDN, focusing on Interactivity

---

**INDCOR WP 4: Evaluation of Interactive Narrative Design for Complexity Representation**
http://arxiv.org/abs/2306.09817
Describes approaches for evaluating the effectiveness of IDNs

---

**INDCOR WP 5: Addressing Societal Issues in Interactive Digital Narratives**
http://arxiv.org/abs/2306.09831
Describes how societal issues are being addressed in IDN artifacts.

# 01. Introduction

Contemporary 21st-century society is facing a multitude of complex interconnected issues and challenges. Interactive Digital Narratives (IDNs) embody and engage such complexity by nature through their expressive and interactive capabilities. IDNs are narrative experiences that can be changed by an audience, and which are created for the digital medium. What can be changed (for example outcome, progression, perspective) varies as does the particular form. Popular IDN forms include narrative-focused video games, interactive documentaries, hypertext fictions and VR/AR/MR experiences. As such, a closer study of IDNs, their users, and their role in addressing social issues is necessary.

The 21st-century social challenges we face are not necessarily novel; nevertheless, they are complexified by structural and historical legacies like coloniality and further imperialism that divides our world between the wealthy North and poor South. Some of these issues include increasing inequality in gender, race, class, and geography, exacerbated by the pandemic, the weakening of democratic states, pressing climate change, other environmental threats, and war. Along with these challenges, citizens around the world have a loss of trust in science and institutions, including those most responsible for informing us about society: our media outlets. At the same time, the complexity of the media landscape, given the exponential rise of social media platforms, has decentralized news outlets in ways that result in forms of incidental consumption and news avoidance, meaning that younger audiences tend to not look for news or are not interested in it.

Our current media and our narrative models that deal with these challenges are no longer effective. This dysfunction, and the resulting chaos, leave us with many critical questions to address: Do we need to radically alter the ways we engage with and study our media and the stories they circulate? Is there a need to create differently to reach new futures or support utopian visions to tackle these issues? How can we foster more hope and care and move from benevolence to solidarity through our shared stories? How do we convey, address, and develop new forms of interactive digital media to address this complexity? How do we foster societal mobilization to go beyond theory and actively promote social change? Addressing these questions and facing these challenges requires a more precise focus on the complexity within and around the narratives we tell and circulate in our media outlets to engage audiences/citizens. In this sense, IDN is a tool to help us to untangle some of these significant issues.

**At their core, IDNs are meant to directly engage with users through their interactive designs and their ability to represent complex issues in an accessible manner.**

These expressive and engaging narratives, designed for digital media delivery, are implemented both as computational systems but, significantly, are meant to be experienced through highly interactive social participation. IDNs are mediated expressions that combines systemic representation and self- guided exploration. The dynamic interplay between the computational systems and their participatory function for users can be approached in different ways and provide a myriad of expressive possibilities and opportunities for application. This means that Interactive Digital Narratives provide ways to represent complex issues and enable understanding through participatory engagement, since audiences can explore them as interactors, making their own choices, picking perspectives, and using the opportunity for replay to revisit earlier decisions and encounter different perspectives.

In this White Paper, we focus on the intricate societal challenges that may provoke a loss of trust in the media to represent reality in its complexity. We recognize there is a deafening effect in how media outlets address societal issues. Recent research shows that social inequalities limit people's access to news as news media outlets increasingly struggle to develop sustainable business models while failing to ensure that the information circulated serves the public good. We ask, then, who is accessing *truthful* content, and how IDNs can alleviate, or not, these intricate problems of access and delivery. How can we leverage participatory practices to address the societal challenges we highlight and engage broader citizenship to address social challenges? We know that the answers to societal complexity and truthfulness within news and media outlets are not straightforward. But we do believe that several of them are already being tackled by new forms of digital media, namely with IDNs, as they circulate around us in an ever-increasing and immersive *life within* media. **In the next subsection, we offer some reflection on how narrative complexity itself both is and has been a core element of societal cohesion and disruption.**

## 01.1. What is Narrative Complexity?

In fact, complexity is a defining feature of human societies. The nature of social complexity is deeply related to two defining, intertwined aspects of sustaining humanity: sociality (i.e., coordination and interdependence among individuals) and symbolic mediation (i.e., the construction of representational systems, such as language and media, that support and facilitate interactions with others and the worlds in which we live).

From a basic organizational perspective, the idea of social complexity involves recognizing interconnectivity among elements within a social order (i.e., people and institutions) and reflexivity— that is, the ability to consider the impact and meaning(s) of these complex interdependent agents and structures in different and changing contexts. **Our challenge is to identify and study such interdependence by focusing on IDNs to better comprehend how they operate within and help form societal complexity across a diverse range of systems.**

We aim to understand the underlying factors that impact the range and intensity of these interconnections and that foster societal complexity, with a particular focus on IDNs as unique systems of representation and interconnection with social systems. We also want to understand how they exist due to society's permanent and enduring effort to deal with its increasing complexity through narrative systems and tools. Digital technologies are a particularly important factor for consideration, having emerged as a resource to cope with complexity deriving from, for example, the pressure of globalization.

Digital technologies can profoundly affect and transform the *communicational tissue* at the core of contemporary societies. They connect us, and they tell us how and what we should know, as well as how we must use that knowledge, for better or worse, to support social structures and systems. In brief, they involve a powerful joining of narrative storytelling capacities with foundational forms for communication to support complex social interconnectivity.

Media forms, as interconnective elements within social systems, have been identified as a kind of societal response to handle levels of complexity, for example, as with industrial societies at the end of the 19th Century that overwhelmed the capacity of other social systems (such as religion, education or politics) to offer effective symbolic representations of society (in novels and in newspaper articles) that could support the common interests and shared decision-making skills necessary among institutions and individuals in a rapidly changing world.

In a sense, media - and particularly news media as it emerged during the 19th and 20th centuries in various forms (newspapers, radio, GV) – constructed a sort of social map that allowed their readers to identify social problems and challenges. In addition, much more importantly, it enabled different social actors to access and navigate that increasingly complex society. Hence, the crucial importance of media—and specifically news media—as a source to recognize how shared narratives allow us to understand ourselves more fully as a society.

IDNs, due to their systemic nature, have the capacity to represent complexity – for example by embodying complex rulesets and thus represent societal, economic, and organic eco systems. On this basis, IDNs effectively introduce multiple elements, events, and characters in their narrative structures, creating a reflexive, participatory, and richly dynamic ground for meaning construction. One especially relevant aspect of IDNs is the focus on the capacity for participatory co-construction of subjects and identities. These capabilities make IDNs an operative device for the understanding of complex issues that can sustain different perspectives, sensibilities and/or emotional experiences concurrently. Also, as IDN actions are built upon the joint premises of participation and interaction, social narratives based on IDN structures may be helpful in the understanding of the causes and consequences of decisions and choices that impact societal contexts. To exemplify the power of IDN to addressing complexity, in the section 2 below, we provide five case studies to better understand how some critical social issues related to the challenge of trusting media representation may be addressed through IDNs. Issues in our case studies focus on several societal challenges: representations of culture and cultural heritage, racism, media disinformation, obesity bias/health awareness, among others.

# 02. Case Studies of IDNs Addressing Societal Issues

## 02.1. Case: Purunmachu, Whispers of the Chachapoyas

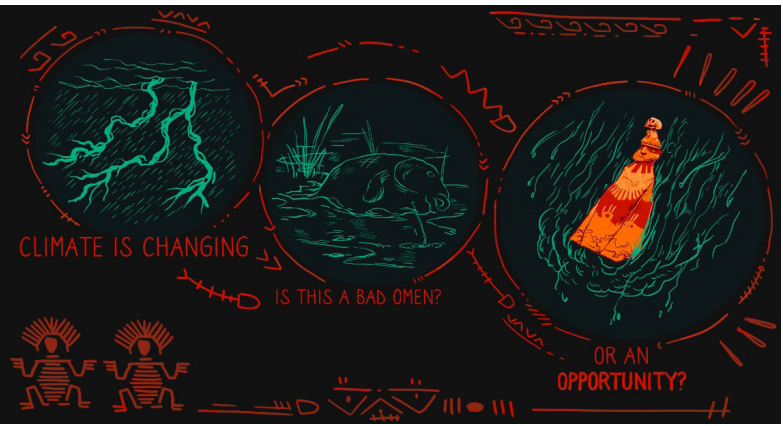

*Figure 1: Screenshot of the game Purunmachu, Whispers of the Chachapoyas.*

### What is this IDN case about?

Purunmachu is an explorer-type game (https://vaniacasu.itch.io/purunmachu) in which you will be able to transport yourself to the Peruvian rainforest, at the edge of the "Sarcófago del Tigre", one important funerary site of the Chachapoyas. There, you'll face your desire. Are you saving a culture? Or helping with the destruction of your own heritage? In the game's climax, the player climbs far above the rainforest canopy and places the Purunmachu among a group of other Purunmachus on the mountaintop. The game ends with the player traveling back home - as told through a comic - and getting some information about the Chachapoyas.

### How does this IDN case reflect societal complexity?

The Chachapoyas was a culture in Peru, believed to be at its peak around 800 AD onward. With their origins in northern Peru – in the Andes – they were also called the Warriors of the Clouds. The culture largely disappeared due to a series of invasions. It went through tumultuous periods with the arrival of the Incas, which required changes in their cultural organization and politics. It was ultimately destroyed during the Spanish colonialism of South America in the 16[th] century. The heritage of the Chachapoyas has, as with many other cultures, been slowly extracted from its origin point. Through colonialism, looting, and careless tourism, many important sites and objects have disappeared.

### Who are the IDN producers?

The development team consisted of two Swedish developers (design and audio), eight Peruvian developers (graphics and writing), and an Iranian developer (programmer). The IDN was produced independently by the team during the Global Game Jam.

### For whom is this IDN produced?

The IDN is currently split between two paths. On one path, the IDN is produced for a general international audience or anyone who might specifically be interested in learning more about the topic and/or those that might find the game's aesthetics appealing. But it is also for those who, through playing the game, hopefully learn more about the subject matter.

On the other path, the focus is on the *community engagement and participatory design* during the development process. In this sense, the IDN is produced more specifically for stakeholders and community members in Peru concerned with critical issues on cultural heritage, representing an opportunity of involving different people in conversations and discussions on the topic.

### Why is this IDN case innovative?

This game is innovative because it offers us a platform to question the social positioning of IDN creators. For example, one might argue that the game ironically can become an act of colonialism and appropriation, because since the game has been developed, when people look for the Chachapoyas - particularly the Purunmachu artifacts - on Google, half of the results are now links to the game jam and pictures of the development team and a conference event. One of the main issues with this type of IDN – a digital game, and especially one developed under the constraints of a game jam – is that it fails to actually give voice to people from the culture it is depicting and over-simplify its message. As such, an IDN like this requires more nuanced development and further refinement and research, but it can also then be a vehicle for learning more about issues and design for interactive cultural heritage representation and for decolonizing histories.

## 02.2. Case: Living Colors: a narrative game about Afro-Resistance

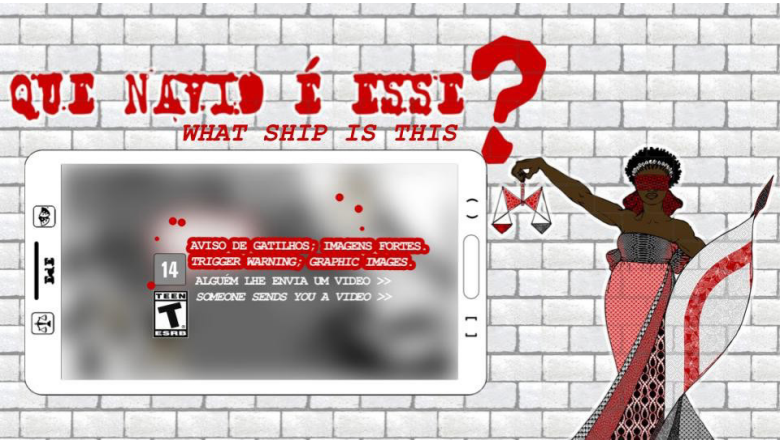

*Figure 2: This is a screenshot of the game Living Colors about Afro Brazilian religion.*

### What is this IDN case about?

This IDN is an online analog role-playing game (https://livingcolors.netlify.app/) that presents the practical and direct consequences of everyday racism. The aim is to provoke a sensation of discomfort to open the space necessary for the manifestation of an Orisha, deities of Yoruba origin revered in Candomblé, an African diasporic religion from Brazil. The game begins with an adventure, *What ship is this?*, to support character creation for players. These characters are Black people who will become messenger spirits of an Orisha after receiving a divine revelation.

### How does this IDN case reflect societal complexity?

The messenger characters are connected to the symbolic and religious repertoires of the African- Diasporic culture, so they cannot manifest in societies that are authoritarian and forcibly homogeneous. In such societies, they tend to be progressively forgotten and risk disappearing. The concept of the Orisha is vital to the game, because the game offers a symbolic alternative to the Judeo-Christian and Greco-Latin imaginaries of character, hero, and identity. From this starting point, the goal is to offer new world views and, at the same time, expose the ethnocentrisms that dominate our culture and circulate within much of our media. The main objective of this IDN is to simulate the experience of building an African-Diasporic identity from the concept of the Orisha.

### Who are the IDN producers?

This IDN was designed by students and professors of the Interactive Stories Research Group and the Decolonial Collective of Black students, both from the Arts and Design Institute of the Federal University of Juiz de Fora (Southeast Brazil). The game follows a *LudoPoetics Method*, which is one of the outcomes of the Ludonarratives & Re-signification Research Line of this research group.

### For whom is this IDN produced?

This IDN was primarily designed for Black people by Black people who share the discomfort of racism daily. That is the discomfort that the Black author [of this case], as well as her non-White students, experience within the contemporary gaming industry and community. This discomfort is intended to be personal and comes from the perspectives of people who know and feel what it is to be on the outskirts of a dominant civilization, and not being represented, included, or heard.

### Why is this IDN case innovative?

This IDN is innovative for questioning Eurocentric standards in narrative games with inspiration from African Brazilian myths and frames of mind to create character concepts and motivations other than the Anglo-German "Hero's Journey" (by Joseph Campbell, for example) and the Eurocentric- archetypal point of view by others like Carl Jung. They support alternative game challenges and solutions other than using narrative models based on *The Good Savior*, a figure used to redeem others from evil, a potentially simplistic and biased perspective based on social and moral constructs.

## 02.3. Case: How RTVE´s Data Journalism Unit Expands Narrative Borders

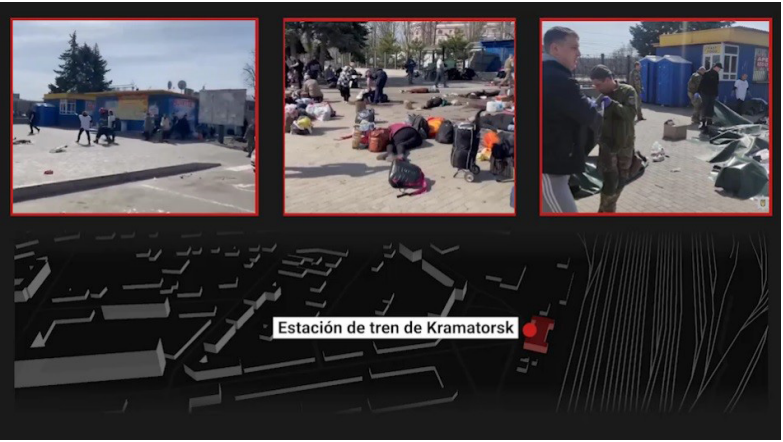

*Figure 3: This is a screenshot of the IDN "War in Ukraine" by RTVE.*

### What is this IDN case about?

This IDN case is the workflow of the RTVE's Data Journalism Unit with all the sections of Spanish public media (RTVE), such as the media lab (RTVE Lab) and the fact-checking unit (Verifica RTVE). In particular, our specific example is the interactive news story/coverage *WAR IN UKRAINE: 15 possible war crimes after 50 days of Russian invasion in Ukraine*, made in collaboration with Verifica RTVE. This IDN contains many elements to support interactivity and digital tools: the contextual use of hyperlinks, data visualization, audiovisual narratives, and external tools such as Google Street View. The high density of multimedia resources offers to the audience an almost endless experience of personalization, and users may choose to design their own experience. This example is available [here](#).

### How does this IDN case reflect societal complexity?

Digital media are the explainers of society for complex issues. Among journalists the concept of the *explainer* can be used to identify a particular piece of content or even a whole medium itself (like Quartz). The Data Journalism Unit, similar to same BBC's Shared Data Unit, helps journalists to find the best narrative means for delivering news. Such ideas are key to supporting Spanish public media and finding best practices. The rise of digital development has disrupted the media landscape by widening information accessibility beyond traditional mainstream media, but also by questioning the quality of the information provided. The result is a new business model that finances quality content for paid subscribers through paywalls or subscriptions.

### Who are the IDN producers?

These IDNs are produced by professionals (journalists, editors, media professionals) in Spain. In particular, the Data Journalism Unit is made up of seven journalists —one of them with a strong profile of specialization on video editing and data visualization-. All texts are available online in Spanish.

### For whom is this IDN produced?

The IDNs are produced for the general public, especially those with interest in the concept of war crimes.

### Why is this IDN case innovative?

Beyond the narrative significance for culture, the ability to rely on the freedom to tell a story with data is critical and requires much research: data, graphs, and other media embedded in news must be critically understood and evaluated. In addition, the multimodadlity of the workflow is an innovative approach to complex societal issues. This Data Journalism Unit supports different sections such as News, Media Lab or Verifica RTVE to maximize the potential of IDNs. Alternately, this approach from the data is useful to understand any coverage as a kind of *precision journalism*.

## 02.4. Case: Disinformation[1] and IDN

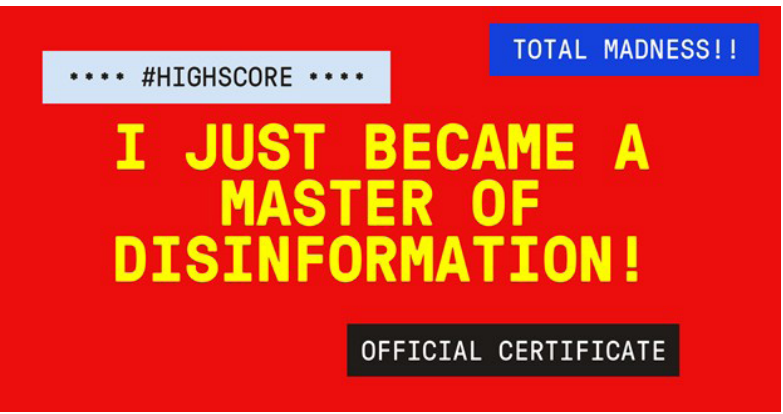

*Figure 4: An official certificate that one gets after playing this game about disinformation.*

### What is this IDN case about?

This IDN, a computer game called "Bad News" ([https://www.getbadnews.com/en](https://www.getbadnews.com/en)) aims to address disinformation and propaganda, by allowing the user to produce different types of disinformation on diverse topics (such as politics, health, climate change, etc.). The aim is to be a 'Master of Disinformation,' and when users achieve that, they learn different tactics of producing and disseminating information disorders (e.g., impersonation, polarization, emotions, conspiracy, trolling, and discredit). At the same time, they are equipped with skills that help them to distinguish and deconstruct propaganda or disinformation.

### How does this IDN case reflect societal complexity?

Disinformation has radically challenged news media hegemony in building social representations. In current information flow dynamics, readers/players/interactors must learn more about disinformation and propaganda, and this game serves this purpose through entertainment and gamification. Disinformation often polarizes society, and this happens also in the game. Thus, it is a direct reflection of societal complexity and the consequences of circulating misinformation.

### Who are the IDN producers?

This IDN is developed by Sander van der Linden, director of the University of Cambridge Social Decision-Making Laboratory, Jon Rozebeek, researcher at the Cambridge Social Decision-Making Lab and DROG, a Netherlands-based platform against disinformation. Later, it was translated by professionals and media practitioners in different countries and now is available in 16 languages: English, Swedish, German, Greek, Polish, Bosnian, Czech, Dutch, Esperanto, Moldovan, Romanian, Serbian, Slovenian, Russian, and Ukrainian.

### For whom is this IDN produced?

The IDN is produced for the general public, aiming through play and fun to improve information seeking people's habits, especially targeting those who enjoy playing digital games.

### Why is this IDN case innovative?

Researchers who developed this game decided on an innovative approach to detect disinformation by implementing the inoculation theory, which suggests that we can strengthen people's resistance to false information by exposing them to weakened versions of that disinformation. The game is choice- based, meaning players are given different options that will affect their journey throughout the game.

Bad News was created for players over the age of 14, and it has a huge player base, surpassing one million players. In this game, players need to attract virtual Twitter followers by falsifying the truth, spreading it, dividing unity, and distracting interest in a busy media context. They also have to stay credible with inside the eyes of their audience. This game extracts truth-destroying techniques into six key strategies: Impersonation, Emotion, Discredit, Trolling, Polarization, and Conspiracy.

---

[1] In this WP, we use the term "disinformation" to refer to content that is intentionally false and designed to cause harm in detriment of "fake news." The latter has lost applicability in academia because it is widely used to attack the press, in a politicization of the term. Besides, "fake news" is considered an oxymoron, as an item cannot be both news and fake at the same time; a piece of content is either news or it is a lie.

## 02.5. Case: 3D computer game titled Nayia's Maze

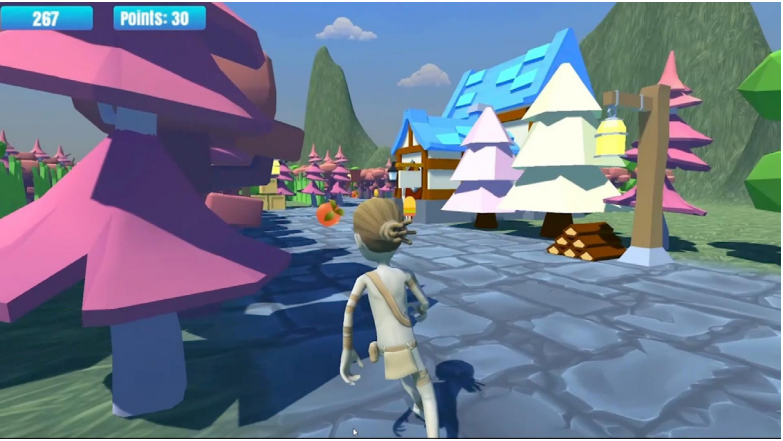

*Figure 5: A screenshot from the IDN "Nayia's Maze" shows the main character running through a maze.*

### What is this IDN case about?

This IDN is a 3D computer game prototype aiming to address obesity as a multifactorial issue while recognizing the importance of good nutrition for children. The main character of the game, a tweenage girl, must move through a maze, overcoming various obstacles, while making decisions concerning food and health within appropriate contexts without succumbing to body stereotyping and stigmatization.

### How does this IDN case reflect societal complexity?

This IDN prototype was intended to represent obesity as a social and health issue to engage students with such a topic through collaboration with peers, jointly creating an IDN with civic significance. The students analyzed research/data addressing the relationship between body weight and media, including framings of obesity in news media, influencers using social media to challenge body-shaming, and literature critiquing media interventions of (childhood) obesity. Disdain toward fatness is not new in societies and media. However, fatness or 'excess' weight, medicalized as an obesity epidemic, has been reframed as a massive, global public health issue and has demonstrated the following tendencies in news media coverage.

### Who are the IDN producers?

This IDN prototype was produced in an educational context by 29 Cypriot communication/animation design students, 16 female and 13 male, who participated in a media literacy intervention as part of the Media Literacy for Living Together (MILT) project funded by the European Commission Directorate-General for Communications Networks, Content and Technology.

### For whom is this IDN produced?

Teenagers, parents, and media educators who prioritize digital media, games, and health literacy.

### Why is this IDN case innovative?

As collaborative-meaning narratives, they are connected to the idea of interaction among users who are re-interpreting the content and the message. The game is an online multi-player game that can be played with friends. Such practices augmented digital literacies for utilizing IDNs in game-based learning and opinion-formation, through which students not only engaged with processes of collaboration and creation but established a sense of civic agency.

# 03. Key Takeaways

Here, we present ten key takeaways concerning the understanding of IDNs as a tool to help us to untangle complex societal issues. They are divided in two broader dimensions, namely: Designing IDNs to **address** such issues and using IDNs to **engage** audiences with them.

Designing IDNs to **address** complex societal issues:

1. The *way* we tell stories is a fundamental part of the way we understand the essential content. Formal design choices must be strategic.
2. Although awareness of other cultures or previously unknown perspectives on critical social issues might be facilitated in IDNs (and other narratives), inspiration for material and social change can only be successful if users are inspired to research the topics independently beyond the original interactive experience. IDN creators need to carefully navigate this issue - otherwise, they risk popularizing simplistic misinterpretations of a subject matter.
3. Technology may be simplified in the process of creating an IDN.
4. The design of choice-based games may be an excellent strategy to develop IDNs that can be used to foster digital-based game learning.
5. Multimodality is a powerful resource for the creation and understanding of IDNs related to complexity.

Using IDNs to **engage** audiences with complex societal issues:

1. Innovation and experimentation with the possibilities for new IDNs are essential if we are to help and engage users in societal contexts to discover new ways to understand information critically.
2. Media outlets cannot renege on their duty of being purveyors of complex societal issues.
3. Showing data visualization according to 'cool' new trends (via Tik Tok, etc,) is not a goal in itself, but a means to reach a wider audience and to support a greater understanding of the social content.
4. Giving freedom to journalists for their choices and processes is key for supporting and growing alternative, counter-traditional, non-hierarchical, and more inclusive workflows.
5. IDN can be used for empowerment and political resistance, and it should be circulated in schools.

# Further Reading

### IDN
- Koenitz, H. (2023). Understanding Interactive Digital Narrative: Immersive Expressions for a Complex Time. Routledge. https://doi.org/10.4324/9781003106425

### News Media and Disinformation
- Boczkowski, P. J., Mitchelstein, E., & Matassi, M. (2018). "News comes across when I'm in a moment of leisure": Understanding the practices of incidental news consumption on social media. New media & society, 20(10), 3523-3539.
- E. & Kligler-Vilenchik, N. (2022). Taking a break from news: a five-nation study of news avoidance in the digital era. Digital Journalism, 10(1), 148-164.
- Gërguri, D. (2023). Kosovo: Frequent Disinformation and a Fertile Ground for Manipulators. In "Blurring the truth: Disinformation in Southeast Europe", edited by Ch. Nehring and H. Sittig, Sofia: KAS Media Programme SEE.
- Qerimi G., & Gërguri D. (2022). Infodemic and the Crisis of Distinguishing Disinformation from Accurate Information: Case Study on the Use of Facebook in Kosovo during COVID-19. Information & Media, 94, 87-109.
- Roozenbeek, J., van der Linden, S. (2019). Fake news game confers psychological resistance against online misinformation. Palgrave Commun 5, 65.
- Wardle, C., & Derakhshan, H. (2018). Thinking about 'information disorder': formats of misinformation, disinformation, and mal-information. Journalism,'fake news'& disinformation, 43-54.

### Decolonial IDN
- Bettocchi, E., Klimick, C., & Perani, L. (2020). Can the Subaltern Game Design? An Exploratory Study About Creating a Decolonial Ludology Framework Through Ludonarratives. In DiGRA '20 – Proceedings of the 2020 DiGRA International Conference: Play Everywhere.
- Mukherjee, S. (2018). Playing Subaltern: Video Games and Postcolonialism. Games and Culture, 13(5), 504-520. https://doi.org/10.1177/1555412015627258
- Reyes, M.C., Silva, C., Koenitz, H. (2023). Decolonizing IDN Pedagogy From and with Global South: A Cross-Cultural Case Study. In: Holloway-Attaway, L., Murray, J.T. (eds) Interactive Storytelling. ICIDS 2023. Lecture Notes in Computer Science, vol 14383. Springer, Cham. https://doi.org/10.1007/978-3-031-47655-6_9
- Silva, C., Reyes, M.C., Koenitz, H. (2022). Towards a Decolonial Framework for IDN. In: Vosmeer, M., Holloway-Attaway, L. (eds) Interactive Storytelling. ICIDS 2022. Lecture Notes in Computer Science, vol 13762. Springer, Cham. https://doi.org/10.1007/978-3-031-22298-6_12

### News Media and IDN
- Koenitz, H. & Louchart, S. (2015). Practicalities and Ideologies, (Re)-Considering the Interactive Digital Narrative Authoring Paradigm. In Li, B., Nelson, M. (eds) Proceeding of the 10th International conference of the Foundations of Digital Games (2015).
- Noguera-Vivo, J.M. (2021). A journalistic approach to IDN when the user is the message: The JUC model for content strategy. h. hal-03226230